\documentclass[onecollarge,runningheads]{svjour2}

\smartqed  

\usepackage{graphicx}

\journalname{Astrophysics and Space Science}

\begin{document}

\title{Subdiffusive transport in
intergranular lanes on the Sun. The Leighton model revisited}
\titlerunning{Subdiffusive transport in
intergranular lanes on the Sun. The Leighton model revisited}

\author{A.A.~Stanislavsky$^1$ and K. Weron$^2$} \authorrunning{A.A.~Stanislavsky and K. Weron}
\institute{$^1$ Institute of Radio Astronomy, 4, Chervonopraporna
St., Kharkov, 61002 Ukraine,\\ \email{alexstan@ri.kharkov.ua}\\
$^2$ Institute of Physics,  Wroc{\l}aw University of Technology,
Wyb. Wyspia$\acute{\rm n}$kiego 27, 50-370 Wroc{\l}aw, Poland,
E-mail: Karina.Weron@pwr.wroc.pl}
\date{Received: date / Accepted: date}

\maketitle

\begin{abstract}
In this paper we consider a random motion of magnetic bright points (MBP)
associated with magnetic fields at the solar photosphere. The MBP
transport in the short time range [0 $-$ 20
minutes]  has a subdiffusive character as the magnetic flux tends
to accumulate at sinks of the flow field.  Such a behavior can be
rigorously described in the framework of a continuous time random walk
leading to the fractional Fokker-Planck dynamics. This formalism, applied
for the analysis of the solar subdiffusion of magnetic fields, generalizes the Leighton's
model.
\end{abstract}

\section{Introduction}

Much of the solar surface phenomena is caused by photospheric
convective motions of magnetic flux ele-ments. Due to a
complicated character of the solar convection, the heliospheric
magnetic field contains a random component. Leighton (1964)
suggested to consider the migration of magnetic regions on the Sun
as a simple Brownian random walk on the solar surface. Until the
precision of experiments was not high, this model was good enough
as an approximation to the diffusion of magnetic elements.
However, the recent experimental studies have shown that the model
is too simple to envelop the phenomenon in full. The MDI
magnetogram data (Hagenaar et al., 1999) from the {\it SOHO}
spacecraft have noticed that the diffusion coefficients in
tracking magnetic elements vary in time. Therefore, Cadavid et al.
(1999) gave a refinement of the Leighton's model. In their
examination the walkers (magnetic elements) stick before the next
jump, i.\ e.\ there are traps at stagnation points slowing down
the walkers' motion. The statistical analysis of the MBP data
approves such a subduffisive transport in the intergranular lanes
within time interval to about 20 minutes. Nevertheless, this point
of view produces an impression of an unfinished work. The
Leighton's model clearly leads to a diffusion equation. It is
hence reasonable to ask whether it is possible to find a similar
equation concerning the subdiffusive case. This question remained
without any answer in the work of Cadavid et al. (1999). Perhaps,
therefore Giacalone and Jokipii (2004) stressed that they do not
see evidence of anomalous diffusion in random walks of magnetic
footpoints. However, one may observe that the analysis of
Giacalone and Jokipii (2004) is related to supergranular scales,
with times being the order of days. Their casual comment about the
results of Cadavid et al. (1999) is somewhat misleading since they
did not investigate the same phenomena. After all, the analysis of
Cadavid et al. (1999) supports the normal diffusion for larger
time scales. Following Simon et al. (1995), the migration of the
magnetic flux from a point to a point tends to accumulate at sinks
of the flow field. The sinks displace randomly. Zimbardo, Veltri
and Pommois (2000) have studied magnetic field line transport in
3D magnetic turbulence with anisotropy in the parallel and
perpendicular directions to the magnetic field. The transport
regime of magnetic field lines depends on parameters like the
magnetic fluctuation level $\delta B/B_0$, the correlation lengths
of magnetic turbulence $l_x, l_y, l_z$ and the dimensionality of
turbulence. The numerical study (Pommois, Veltri and Zimbardo,
1999) shows that the transport can be anomalous (subdiffusive or
superdiffusive). The various transport regimes are conveniently
classified in terms of the Kubo number (Pommois, Veltri and
Zimbardo, 2001). For magnetic turbulence this number is defined as
$R=(\delta B/B_0)(l_{||}/l_\bot)$ , where $l_{||}=l_z$ is the
correlation length parallel to the average field
$\vec{B_0}=B_0\vec{e}_z$, $l_\bot=l_x=l_y$ the correlation length
perpendicular to $\vec{B_0}$. For $R\leq 0.2$ there are anomalous
non-Gaussian transport regimes, whereas for $0.2<R\leq 1$ there is
an approximately quasilinear Gaussian diffusive regime. From this
point of view the analysis of macroscopic diffusion equations for
the motion of magnetic field lines represents a great interest in
physics and astrophysics.

Recent progress (Metzler and Klafter, 2000; Zaslavsky, 2002;
Stanislavsky, 2004; Meerschaert and Scheffler, 2004; Magdziarz et
al., 2007; Magdziarz and  Weron 2007) in understanding anomalous
diffusion allows one to represent a subdiffusive transport of
magnetic fragments on the solar surface in a more comprehensive
form. The purpose of this paper is to perform this work. A simple
mathematical review on the random processes of anomalous diffusion
is given in Section~2. This view will be especially useful for
readers far from the probabilistic theory. In Section~3 we derive
the fractional Fokker-Planck equation (FPE) in spherical
coordinates. This particular equation corresponds to a rotational
subdiffusion being of interest to the solar physics of magnetic
fields. In next section we discuss the solution describing the
migration of the MBP on the Sun. Section~5 contains conclusions.

\section{Diffusive processes as a continuous limit of CTRW}

The notion of continuous time random walks (CTRW) has been
introduced in physics by Montroll and Weiss (1965). They
generalized a simple random walk which is based on the assumption
that step changes (jumps) are made through equal time intervals.
In contrast, the CTRW concerns random walks with a random waiting
time among subsequent random jumps. The generalization has become
very popular for many physical applications of anomalous
diffusion, e.\ g. for transport in disordered media, superslow
relaxation, etc. However, this model is useless until random
values (jumps and waiting times) are undefined by a probabilistic
description (for example, by their probability densities or
characteristic functions).

In order to explain the idea of a CTRW let us consider the simplest one-dimensional case.
The  position of a walker after $k$ random jumps is given by
\begin{equation}
R(k)=\sum^k_{i=1}R_i, \quad R(0)=0\,,\label{eq1}
\end{equation}
where $\{R_i\}$ are random variables representing the length and
the direction of jumps. The corresponding time interval reads
\begin{equation}
T(k)=\sum^k_{i=1}T_i, \quad T(0)=0\,,\label{eq2}
\end{equation}
where $\{T_i\}$ represent the random time intervals (waiting time)
between successive jumps of a walker. The number $N_t$ of jumps
performed by the walker till time $t$ is a counting process given
by the following relation
\begin{equation}
N_t={\rm max}\{k:T(k)\leq t\}\,,\label{eq3}
\end{equation}
meaning that the random number $N_t$ of jumps occurred up to time
$t$ is equal to the largest index $k$ for which the sum
$T_1+T_2+\dots T_k=T(k)$ of $k$ random time intervals does not
exceed time $t$. Consequently, the total distance reached by the
walker up to time $t$ determines the stochastic process
\begin{equation}
R(N_t)=\sum^{N_t}_{i=1}R_i\,,\label{eq4}
\end{equation}
known as the CTRW. The analysis of probabilistic properties of
such a random sum as given in (\ref{eq4}) is a core of limit
theorems in probability theory.  Limit theorem, under a certain
necessary and sufficient mathematical conditions, yields the
continuous limit of the sums (or any other operation on the
sequences of random variables , following Feller, 1971). In
particular, taking into account the expected values $\langle
R_i\rangle<\infty$  and $\langle T_i\rangle\to\infty$, we get the
continuous limit $X(S_t)$ of the CTRW process (\ref{eq1}). Let us
note that the infinite expected value of the waiting time $T_i$ is
connected with a long-tail property of its probability density
function: $h(t)\sim t^{-\alpha-1}$ for $t\to\infty$ and
$0<\alpha<1$ (for more details, see Magdziarz and Weron, 2006).
The limiting  process, known as the anomalous diffusion process,
is expressed in terms of the parent process $X(\tau)$ subordinated
by the random time clock $S_t$. In terms of probability density
functions we can write the probability density of the subordinated
process $X(S_t)$ as an integral relation
\begin{equation}
p(x,t)=\int^\infty_0 f(x,\tau)\,g(t,\tau)\,d\tau\,,\label{eq5}
\end{equation}
where $f(x,\tau)$ and $g(t,\tau)$ are the probability density
functions of $X$ and $S_t$, respectively. Here, the probability
density $f(x,\tau)$ represents the probability of finding the
parent process $X(\tau)$ at $x$ on the operational time $\tau$,
whereas $g(t,\tau)$ describes the probability for the operational
time $\tau$ to coincide with the real time $t$. The physical
interpretation of  $S_t$ is that this process accounts for the
amount of time, when a walker does not participate in motion
(Baeumer et al., 2005). If the walker randomly moves all time, the
operational time coincides with the physical one, and $g(t,\tau)$
is simply the Dirac $\delta$-function.

It is very important that the functional form of the probability
density of the random variable $S_t$ can be calculated explicitly.
The procedure can be given in few steps. According to Bingham
(1971), the Laplace transform of the probability density
$g(t,\tau)$ with respect to $\tau$ equals
\begin{displaymath}
\bar{g}(t,v)=\int^\infty_0e^{-v\tau}\,g(t,\tau)\,d\tau=\langle
e^{-vS_t}\rangle=E_\alpha(-vt^\alpha)\,,
\end{displaymath}
where  $E_\alpha(-vt^\alpha)$ is the Mittag-Leffler function. The
Mittag-Leffler function has a simple Laplace image  with respect
to $t$, namely
\begin{displaymath}
\int^\infty_0e^{-ut}\,\bar g(t,v)\,dt=u^{\alpha-1}/(u^\alpha+v).
\end{displaymath}
The latter expression is easily inverted by Laplace with respect
to $v$ as an exponential function. As a consequence, the Laplace
image of $g(t,\tau)$ with respect to $t$ is
\begin{displaymath}
\tilde{g}(u,\tau)=\int^\infty_0e^{-ut}\,g(t,\tau)\,dt=u^{\alpha-1}
\exp\{-u^\alpha\tau\}.
\end{displaymath}
By taking now the inverse Laplace transform of $\tilde{g}(u,\tau)$ with
respect to $u$, we obtain the probability density of the process
$S_t$ in the form
\begin{equation}
g(t,\tau)=t^{-\alpha}F_\alpha(\tau/t^\alpha)\,, \label{eq6}
\end{equation}
where the function $F_\alpha(z)$ can be written as a Taylor series
\begin{displaymath}
F_\alpha(z)=\sum_{n=0}^\infty\frac{(-z)^n}{n!
\Gamma(1-\alpha-n\alpha)}\,.
\end{displaymath}
Thus, according to Eq. (\ref{eq5}), the probability density
$p(x,t)$ reads
\begin{equation}
p(x,t)= \int^\infty_0F_\alpha(z)\,
f(x,t^\alpha z)\,dz\label{eq7}
\end{equation}
and its Laplace image takes the form
\begin{equation}
\tilde{p}(x,u)=\int_0^\infty
p(x,t)\,\exp\{-ut\}\,dt=u^{\alpha-1}\tilde{f}(x,u^\alpha),\label{eq7a}
\end{equation}
where
\begin{displaymath}
\tilde{f}(x,u^\alpha)=\int_0^\infty
f(x,\tau)\,\exp\{-u^\alpha\tau\}\,d\tau.
\end{displaymath}
The representation (\ref{eq7a}) will be used in the next section for derivation
of the macroscopic equation of anomalous diffusion.

The above-mentioned mathematical techniques, appropriate for
translational motion of a walker, can be also applied for a
rotational random walk. As for the motion of MBP, associated with
magnetic fields at the photosphere, subdiffusion on a sphere is of
interest to us. This is nothing else but a rotational
subdiffusion. Then the space jumps will be given by a polar angle
$\theta$ and a longitude $\phi$ in the spherical coordinates. It
is important to observe that the density $f(x,\tau)$ may be
governed by a Fokker-Plank equation (FPE) with a time-independent
potential, discussed in the next section.

\section{Equation of rotational subdiffusion}
To derive an equation of rotational subdiffusion, we start with
one-dimensional case. Let $\hat L(x)$ be a time-independent
Fokker-Planck operator well-known in the classical statistical
physics. Assume that the probability density $f(x,\tau)$ describes
an ordinary Brownian motion with respect to the operational time
$\tau$. It will satisfy the FPE with the operational time:
\begin{displaymath}
\partial f(x,\tau)/\partial\tau=\hat
L(x)\,f(x,\tau)\,.
\end{displaymath}
Applying operator $\hat L(x)$ to the Laplace image (\ref{eq7a}),
we find the following expression
\begin{displaymath}
\hat L(x)\,\tilde{p}(x,u)=
u^\alpha\,\tilde{p}(x,u)-q(x)\,u^{\alpha-1},
\end{displaymath}
where $q(x)$ is the initial condition. The inverse Laplace
transform of the latter gives the fractional integral form of the
FPE
\begin{equation}
p(x,t)=q(x)+\frac{1}{\Gamma(\alpha)}\int_0^td\tau
(t-\tau)^{\alpha-1}\hat
L(x)\,p(x,\tau)\,.\label{eq8}
\end{equation}
The kernel of this integral is a power function resulting
in long-term memory effects for the process $X(S_t)$.
This memory is a direct consequence of subordination of the
space variable $X(\tau)$. For $\alpha=1$, equation (\ref{eq8})
reduces to the ordinary FPE without any memory effects. It
should be mentioned that the
integral form of equation (\ref{eq8}) corresponds to the
equivalent differential form
\begin{displaymath}
\frac{\partial^\alpha p(x, t)}{\partial
t^\alpha}-\frac{q(x)\,t^{-\alpha}}{\Gamma(1-\alpha)}=\hat
L(x)\,p(x,t)\,,
\end{displaymath}
where $\partial^\alpha/\partial t^\alpha$ denotes the
Liouville-Riemann fractional differential operator of order
$\alpha$ (Samko et al., 1993). The Liouville-Riemann
fractional differential operator of integer order is simply
the ordinary derivative.

As a consequence, the equation of rotational subdiffusion in the
simplest case of motion on a circle reads
\begin{equation}
\frac{\partial^\alpha W_\alpha(\theta,t)}{\partial
t^\alpha}-\frac{W_\alpha(\theta,0)\,t^{-\alpha}}{\Gamma(1-\alpha)}=
\frac{D_\theta}{\sin\theta}\,
\frac{\partial}{\partial\theta}\left(\sin\theta\,\frac{\partial
W_\alpha(\theta,t)}{\partial\theta}\right)\,,\label{eq8a}
\end{equation}
where $D_\theta$ is the diffusion coefficient and
$W_\alpha(\theta,0)$ the initial condition. Here, the probability
density $W(\theta,t)$ depends on the polar angle $\theta$ only.
The solution of Eq.(\ref{eq8a}) may be expressed in terms of the
following integral transformation
\begin{displaymath}
W_\alpha(\theta,t)=\int^\infty_0F_\alpha(z)\, W_1(\theta,t^\alpha
z)\,dz\,.
\end{displaymath}
This implies that the averaged value of $\cos\theta$ equals
\begin{displaymath}
\langle\cos\theta\rangle=\int_{0}^{\pi}\cos\theta\,
W_\alpha(\theta,t)\,\sin\theta\,d\theta=E_\alpha(-2D_\theta
t^\alpha)\,.
\end{displaymath}
For large $t$ all the directions of the walker motion become
equiprobable. However, in contrast to the normal rotational
diffusion this equilibrium is reached slowly because of a power
asymptotics of the Mittag-Leffler function.

\section{Transverse subdiffusion of magnetic footpoints at the Sun}

We now write down the equation of subdiffusion associated with the
observed motion of magnetic footpoints embedded in the transverse
flows on the solar surface. The distribution $g_\alpha(\theta,
\phi\,, t)$ of magnetic field footpoints on the solar spherical
surface of radius $R_\odot$ obeys the two-dimensional equation in
spherical coordinates
\begin{equation}
\frac{\partial^\alpha g_\alpha}{\partial
t^\alpha}-\frac{g_\alpha(\theta,\phi,0)\,t^{-\alpha}}{\Gamma(1-\alpha)}=
\frac{1}{R_\odot\sin\theta}\,
\frac{\partial}{\partial\theta}\left(\frac{\kappa\sin\theta}{R_\odot}\,\frac{\partial
g_\alpha}{\partial\theta}\right)+\frac{1}{R_\odot\sin\theta}\,\frac{\partial}
{\partial\phi}\left(\frac{\kappa}{R_\odot\sin\theta}\,\frac{\partial
g_\alpha}{\partial\phi}\right)\,,\label{eq9}
\end{equation}
where $\kappa$ is the subdiffussion coefficient with the following
dimension $[\kappa]=1/$time\,$^\alpha$. The solution of Eq.
(\ref{eq9}) can be obtained via a separation ansatz in terms of
the spherical harmonics $Y_{mn}(\theta,\phi)$, namely
\begin{equation}
g_\alpha(\theta,\phi,t)=
\sum^\infty_{n=0}\sum_{m=-n}^{m=n}a_{mn}\,Y_{mn}(\theta,\phi)\,
E_\alpha\left(-\frac{\kappa}{R^2_\odot}\,n(n+1)\,t^\alpha\right)\,.\label{eq10}
\end{equation}
If one takes an impulse release of footpoints at the location
$(\theta_0,\phi_0)$, the initial condition is determined through
the Dirac $\delta$-function
\begin{displaymath}
g_\alpha(\theta,\phi,0)=\frac{1}{\sin\theta}\,\delta(\phi-\phi_0)\,
\delta(\theta-\theta_0).
\end{displaymath}
Using the completeness relationship of spherical harmonics states, one obtains
the amplitude coefficients $a_{mn}$ expressed by the
complex conjugate of the spherical harmonics
$Y^*_{mn}(\theta_0,\phi_0)$. It is interesting to consider an
asymptotic behavior of the solution
(\ref{eq10}) as $t\to\infty$. The distribution approaches asymptotically the limit
\begin{displaymath}
\lim_{t\to\infty}g_\alpha(\theta,\phi,t)=\frac{1}{4\pi},
\end{displaymath}
which simply corresponds to
the homogeneous distribution on the sphere in equilibrium.
After integrating expression (\ref{eq10}) from $\phi=0$ to
$2\pi$, we get the solution of Eq. (\ref{eq9}) which  will depend only on the
polar angle $\theta$ and time $t$:
\begin{equation}
g_\alpha(\theta,t)=\sum^\infty_{n=0}\frac{2n+1}{2}\,P_n(\cos\theta_0)\,P_n(\cos\theta)\,
E_\alpha\left(-\frac{\kappa}{R^2_\odot}\,n(n+1)\,t^\alpha\right),\label{eq11}
\end{equation}
where $P_n(y)$ is the Legendre polynomial of degree $n$. All the
odd moments of the distribution are equal to zero, but the even
moments are not. So the first even moment takes the form
\begin{displaymath}
\langle\sin^2\theta\rangle=\frac{2}{3}\left(1-E_\alpha\Bigl(-\,\frac{6\kappa}{R^2_\odot}\,
t^\alpha\Bigr)\right)\,.
\end{displaymath}
Since the subdiffusion motion of magnetic footpoints is
characterized by a small value $\kappa/R^2_\odot$ (about $ 0.001$
in order of magnitude), we arrive at the relation
\begin{equation}
\langle\theta^2\rangle\approx\frac{4\kappa}{\Gamma(1+\alpha)R^2_\odot}\,
t^\alpha\,.\label{eq12}
\end{equation}
It is obvious that for $\alpha=1$ this description leads to the
Leighton's model.

In the framework of the fractional FPE approach to the transport
of magnetic fields on the Sun we are able to recover the
experimental evidence which demonstrates the change of diffusion
properties during the life time of the MBP, from subdiffusion to
normal one. From the theoretical point of view the difference
between subdiffusion and ordinary diffusion is related with an
evolution of the value of the parameter $\alpha$, from $\alpha<1$
to $\alpha=1$. This change corresponds to different properties of
the distribution of the interjump time intervals of the MBPs. If
the expected value of the waiting time is infinite ($0<\alpha<1$),
then bright points have to be trapped. This effect is connected
with the long-tailed properties of the waiting-time distribution,
the necessary condition to obtain the subdiffusion from the CTRW
scheme. For $\alpha$=1 there are no traps, since the expected
waiting-time value is finite or time variable is deterministic.
Due to the traps the diffusion of the MBPs has a mixed character
of random stops and motion,  whereas in the case of normal
diffusion the motion of the MBPs continues all time. From the
astrophysical point of view the appearance of subdiffusion in the
short-time range means that the life time of the traps is shorter
than the life line of the MBP. Following the magnetic turbulence
studies on laboratory and astrophysical plasmas, the anomalous
diffusion of magnetic field lines may be associated with the
existence of closed magnetic surfaces. Probably, the surfaces
serve as traps for the MBP. The slope of the experimental data
variance in time is a criterion for revealing the peculiarity. The
comparison of the subdiffusive model with experimental data may be
carried out in the same manner as this is the case for the work of
Cadavid et al. (1999).

\section{Conclusions}
We have presented an approach to anomalous diffusion which follows
from an intuitive concept of sticking times suggested in the paper
of Cadavid et al. (1999). We have derived the fractional FPE in
spherical coordinates appropriate for describing the subdiffusive
migration of the MBP. As a special case, the normal diffusion of
the MBPs on the Sun can be obtained. Therefore, the Leighton's
model has been generalized.  It should be noticed that the
consideration of two different diffusive regimes in the MBP
motion, as resulting from a simple sum of an ordinary first
temporal derivative and a fractional temporal one in the FPE, will
not lead to the expected phenomenon. In this case for short-time
region such a temporal operator would lead to a normal diffusion,
whereas for the large-time scale it does a subdiffusion (see, as
an example, the work of Schumer et al., 2003).

\begin{acknowledgements}
The work was partly realized within the framework of the project
INTAS-03-5727. A.A.S. is grateful to A. C. Cadavid for
fruitful remarks.
\end{acknowledgements}

\end{document}